# Precise observations of the $^{12}$C/$^{13}$C ratios of HC$_3$N in the low-mass star-forming region L1527

Mitsunori Araki[1,2], Shuro Takano[3], Nami Sakai[4], Satoshi Yamamoto[5], Takahiro Oyama[1], Nobuhiko Kuze[6], and Koichi Tsukiyama[1,2]

[1] Department of Chemistry, Faculty of Science Division I, Tokyo University of Science, 1-3 Kagurazaka, Shinjuku-ku, Tokyo 162-8601, Japan; araki@rs.kagu.tus.ac.jp
[2] IR Free Electron Laser Research Center, Research Institute for Science and Technology (RIST), Tokyo University of Science, 2641, Yamazaki, Noda, Chiba 278-8510, Japan
[3] Department of Physics, General Studies, College of Engineering, Nihon University, 1 Nakagawara, Tokusada, Tamuramachi, Koriyama, Fukushima 963-8642, Japan
[4] The Institute of Physical and Chemical Research (RIKEN), 2-1 Hirosawa, Wako, Saitama 351-0198, Japan
[5] Department of Physics and Research Center for the Early Universe, The University of Tokyo, Bunkyo-ku, Tokyo 113-0033, Japan
[6] Department of Materials and Life Sciences, Faculty of Science and Technology, Sophia University, 7-1 Kioi-cho, Chiyoda-ku, Tokyo 102-8554, Japan

## ABSTRACT

Using the Green Bank 100 m telescope and the Nobeyama 45 m telescope, we have observed the rotational emission lines of the three $^{13}$C isotopic species of HC$_3$N in the 3 and 7 mm bands toward the low-mass star-forming region L1527 in order to explore their anomalous $^{12}$C/$^{13}$C ratios. The column densities of the $^{13}$C isotopic species are derived from the intensities of the $J$ = 5–4 lines observed at high signal-to-noise ratios. The abundance ratios are determined to be 1.00:1.01 ± 0.02:1.35 ± 0.03:86.4 ± 1.6 for [H$^{13}$CCCN]:[HC$^{13}$CCN]:[HCC$^{13}$CN]:[HCCCN], where the errors represent one standard deviation. The ratios are very similar to those reported for the starless cloud, Taurus Molecular Cloud-1 Cyanopolyyne Peak (TMC-1 CP). These ratios cannot be explained by thermal equilibrium, but likely reflect the production pathways of this molecule. We have shown the equality of the abundances of H$^{13}$CCCN and HC$^{13}$CCN at a high-confidence level, which supports the production pathways of HC$_3$N *via* C$_2$H$_2$ and C$_2$H$_2$$^+$. The average $^{12}$C/$^{13}$C ratio for HC$_3$N is 77 ± 4, which may be only slightly higher than the elemental $^{12}$C/$^{13}$C ratio. Dilution of the $^{13}$C isotope in HC$_3$N is not as significant as that in CCH or c-C$_3$H$_2$. We have also simultaneously observed the DCCCN and HCCC$^{15}$N lines and derived the isotope ratios: [DCCCN]/[HCCCN] = 0.0370 ± 0.0007 and [HCCCN]/[HCCC$^{15}$N] = 338 ± 12.

*Subject Keywords*: Astrochemistry—ISM: clouds—ISM: molecules—ISM: individual (L1527)—Radio lines: ISM






# 1. INTRODUCTION

Carbon-chain molecules constitute 40% of the molecular species detected in interstellar space. They are known to be abundant in cold starless cores in early chemical evolutionary stages (Suzuki et al. 1992). Using this chemical behavior, carbon-chain molecules have been extensively employed as good tracers of the early phase of prestellar core evolution (Benson et al. 1998; Aikawa et al. 2001; Rathborne et al. 2008; Hirota et al. 2009; Marka et al. 2012). Carbon-chain molecules are also known to be abundant in a dense and warm region of some low-mass protostellar cores. This is called "warm carbon-chain chemistry (WCCC)," and its prototypical source is L1527 in Taurus (e.g., Sakai et al. 2008, 2009a; Sakai & Yamamoto 2013).

General features of the two types of carbon-chain chemistry are successfully explained by chemical models (e.g., Aikawa et al. 2001, 2008; Hassel et al. 2008). However, production pathways of individual molecular species are often controversial and are still subject to detailed investigation. For instance, the various $^{12}C/^{13}C$ ratios for nonequivalent carbon atoms constituting a carbon-chain molecule have posed an important problem in astrochemistry. Anomalies in the $^{12}C/^{13}C$ ratio are reported for $HC_3N$ (Takano et al. 1998, Li et al. 2016, Taniguchi et al. 2016a), CCS (Sakai et al. 2007), CCH (Saleck et al. 1994, Sakai et al. 2010b), $C_3S$ (Sakai et al. 2013), $C_4H$ (Sakai et al. 2013), cyclic-$C_3H_2$ (Yoshida et al. 2015), and $HC_5N$ (Takano et al. 1998, Taniguchi et al. 2016b) and can be discussed in terms of their respective formation pathways and some exchange reactions after formation.

The $^{13}C$ isotope anomaly of $HC_3N$ in the starless cloud Taurus Molecular Cloud-1 Cyanopolyyne Peak (hereafter referred to as TMC-1 CP) was investigated by Takano et al. (1998). The column densities of $H^{13}CCCN$, $HC^{13}CCN$, and $HCC^{13}CN$ were reported to be $(2.0 \pm 0.2) \times 10^{12}$, $(2.1 \pm 0.2) \times 10^{12}$, and $(2.9 \pm 0.3) \times 10^{12}$ cm$^{-2}$, respectively, where the errors denote one standard deviation. Essentially, these abundance ratios reduce to 1.0:1.0:1.4 for $[H^{13}CCCN]:[HC^{13}CCN]:[HCC^{13}CN]$. On the basis of these ratios, Takano et al. (1998) pointed out that the reaction of $C_2H_2 + CN$ (Herbst & Leung 1990, Fukuzawa & Osamura 1997, Woon & Herbst 1997) is the most important pathway because the abundances of $[H^{13}CCCN]$ and $[HC^{13}CCN]$ should be equal to each other if the triple bond of $C_2H_2$ is conserved without breaking in the reaction. In addition, the reactions of $C_2H_2^+ + HCN$ (Huntress 1977, Schiff & Bohme 1979) and $C_2H_2 + HCNH^+$ (Mitchell et al. 1979) can also contribute to $HC_3N$ production. Thus, it was suggested that $HC_3N$ can be produced by reactions *via* $C_2H_2$ or $C_2H_2^+$ having two equivalent carbon atoms.

This pioneering result opened a new avenue for studying microscopic chemical processes in interstellar clouds through observations of $^{13}C$ isotopic species. However, the above argument is





based on the equal abundances of H$^{13}$CCCN and HC$^{13}$CCN. Although the abundances of the two species are similar to each other in TMC-1 CP, their ratio still has a mutual error of 14% in one standard deviation as the $J$ = 4–3 rotational lines reported by Takano et al. (1998) have a signal-to-noise (S/N) ratio of 27 at most. The lines of the $^{13}$C species of HC$_3$N in L1527 were recently observed by Taniguchi et al. (2016a), in which the abundance ratios of the three $^{13}$C isotopic species of HC$_3$N were reported to be 0.9 ± 0.2 : 1.00 : 1.29 ± 0.19 (one standard deviation) for [H$^{13}$CCCN]:[HC$^{13}$CCN]:[HCC$^{13}$CN] based on the $J$ = 9–8 and 10–9 rotational lines having an S/N ratio of 11 at most. This observational result in L1527 implies equal abundances of H$^{13}$CCCN and HC$^{13}$CCN, but the ratios have large errors. It is therefore important to confirm the equal abundances of these two species as precisely as possible in order to stringently constrain the formation processes. Furthermore, the possible contribution of reactions exchanging the position of the $^{13}$C atom within a molecule has been proposed for other carbon-chain molecules (Sakai et al. 2010b; Furuya et al. 2011), which is a possibility that needs to be examined carefully.

With these issues in mind, we observed the rotational transitions of the normal and isotopic species of HC$_3$N in order to precisely derive the isotopic abundance ratios in the low-mass star-forming region L1527, where HC$_3$N has been found to be abundant (Sakai et al. 2009b). The rotational spectral lines of DCCCN, HCCC$^{15}$N, and H$^{13}$CCCCCN were also observed simultaneously. We discuss HC$_3$N production schemes in L1527 using these observational results.

## 2. OBSERVATIONS

### 2.1. Observations with the Green Bank 100 m Telescope

The $J$ = 5–4 rotational lines of the three $^{13}$C isotopic species and the normal species of HC$_3$N in the 42.2–45.5 GHz region listed in Table 1 were observed simultaneously using the Robert C. Byrd Green Bank Telescope (GBT) of the National Radio Astronomy Observatory[1] from March to October 2015. We observed the IRAS 04368+2557 position in L1527: ($\alpha$2000.0, $\delta$2000.0) = (04$^h$ 39 $^m$ 53$^s$.89, 26°03′11.0″) (Sakai et al. 2008). The dual polarization Q-band receiver, whose instantaneous bandwidth is 4 GHz, was used. The system temperature during the observations ranged from 70 to 100 K depending on the elevation angle of the telescope and weather conditions. The main-beam efficiency is 0.83 and 0.81 in the 42.2 and 44.1–45.5 GHz regions, respectively (GBT Support Staff 2014). The half-power beam width (HPBW) of the telescope is 16.6″ at 45.5 GHz. The pointing of the telescope was checked by observing the nearby continuum sources,

---

[1] The National Radio Astronomy Observatory is a facility of the National Science Foundation operated under cooperative agreement by Associated Universities, Inc.





J0336+3218 and J0319+4130, every 1.5 h, leading to a pointing accuracy of 5″. The VErsatile GBT Astronomical Spectrometer (VEGAS) was used as a backend in the 23.44-MHz bandwidth and 1.4-kHz resolution mode, corresponding to a velocity resolution of 0.15–0.16 km/s by 16-channel smoothing. A frequency-switching mode with a frequency offset of ±2.5 MHz was employed for the observations. The intensity scale was calibrated by noise injection. To obtain the final spectra, the right- and left-circular polarization spectra were averaged.

### *2.2. Observations with the Nobeyama 45 m telescope*

Observations in the 81.5–109.2 GHz region were carried out in 2008–2010 using the Nobeyama Radio Observatory (NRO) 45 m telescope,[2] where the 88.2-109.2 GHz region was observed as part of the Nobeyama 45 m Telescope Legacy Project. We observed the same position observed in the GBT measurements above. The dual-polarization side-band separating SIS mixer receivers T100V/H were used in the 88.2–91.0 and 109.2 GHz regions in which the spectra obtained using T100V and T100H were averaged. For observations in the 81.5 GHz region, only the T100V receiver was used. The main-beam efficiency is 0.49 and 0.42 in the 81.5 and 88.2–109.2 GHz regions, respectively. The beam size of the telescope is 19.6″, 18.1″, 17.7″, and 15.5″ at 81.5, 88.2, 91.0, and 109.2 GHz, respectively. The telescope pointing was checked by observing the nearby SiO maser ($v = 1$, $J = 1$–0) in NML Tau every 1.5 h. The typical pointing accuracy was a few arc-seconds. A position switching mode with the off position taken at $\Delta\alpha = 30′$ and $\Delta\delta = 30′$ was employed for the observations. A set of acousto-optical spectrometers (AOSs) was used for the backend. The bandwidth and the frequency resolution of each wide-band AOS are 250 MHz and 250 kHz, respectively, and those of each high-resolution AOS are 40 MHz and 37 kHz, respectively. The frequency resolutions of the wide-band and high-resolution AOSs correspond to velocity resolutions of 0.92 km s$^{-1}$ and 0.14 km s$^{-1}$, respectively, at 81.5 GHz. The intensity scale was calibrated using a chopper wheel method.

### **3. RESULTS**
### *3.1. Column Density*

The observed lines of the isotopic and normal species of HC$_3$N are shown in Figures 1–3, and their line parameters obtained from Gaussian fitting are listed in Table 1. Due to the high sensitivity of the GBT, S/N ratios as high as 35–44 were achieved for the $J = 5$–4 lines of the three

---

[2] The 45 m telescope is operated by the Nobeyama Radio Observatory, a branch of National Astronomical Observatory of Japan.





$^{13}$C isotopic species, which are less abundant and consequently have weak lines ($T_{MB} \leq 0.1$ K). Two weak satellite hyperfine components of the $J = 5-4$ transition for the normal species were also detected with S/N = 53 and 54, respectively.

As shown in Figure 1, the $J = 5-4$ line for HCC$^{13}$CN is clearly brighter than those for H$^{13}$CCCN and HC$^{13}$CCN. This trend can be confirmed through a comparison of the integrated intensities (Table 1). A similar trend is also seen in the $J = 10-9$ lines, although the S/N ratio is not as high as for the $J = 5-4$ lines. Thus, a $^{12}$C/$^{13}$C anomaly is evident among the three carbon atoms of HC$_3$N in L1527.

To enable quantitative analyses of the $^{12}$C/$^{13}$C anomaly, the column density and the excitation temperature of each $^{13}$C isotopic species of HC$_3$N are determined through least-square fitting of the integrated intensities of the observed lines under the assumption of local thermodynamic equilibrium (LTE). The $J = 9-8$ lines of HC$^{13}$CCN and HCC$^{13}$CN are half-weighted in the least-square fitting because they have lower S/N ratios and lower frequency resolution. Here, the source coupling factor is assumed to be unity, which is justified because the C$_4$H and CCH emitting regions of 30–40″ (Sakai et al. 2008, 2010a) are larger than the beam sizes of the Green Bank and NRO observations. Partition functions are numerically calculated by summing over rotational energy levels up to $J = 60$. The dipole moment of $\mu = $ **3.73** D for the ground vibrational state of HCCCN (DeLeon & Muenter 1985) is also used for all of the $^{13}$C isotopic species. Results of the least-squares fit are summarized in Table 2. The excitation temperatures obtained for the three species H$^{13}$CCCN, HC$^{13}$CCN, and HCC$^{13}$CN are similar. The maximum optical depth is 0.01 for the $J = 5-4$ lines of HCC$^{13}$CN.

In this study, the effects of the hyperfine splitting owing to the $^{14}$N nucleus ($I = 1$) are treated as follows. We use unsplitted intensities without hyperfine structures in the evaluation of column densities and excitation temperatures. The observed lines of HC$_3$N and its isotopic species (except for HCCC$^{15}$N) consist of the three main hyperfine components ($\Delta F = +1$) in the $J = (J'' + 1) - J''$ rotational transition, which are blended together. In addition, the two satellite components ($\Delta F = 0$) appear at both sides of the main components. As the satellite components are much weaker than the main components, they can be detected only for the normal species. However, their contributions need to be considered in order to evaluate the unsplitted intensity without the hyperfine structure. The intensities of the three strong hyperfine components constitute 97.3, 99.2, 99.3, and 99.5% of the unsplitted intensities of the $J = 5-4$, 9–8, 10–9, and 12–11 transitions, respectively (Townes & Schawlow 1955). To obtain column densities and excitation temperatures, the integrated intensities corrected by these factors are used and are listed as $W_{all}$ in Table 1.

We accurately derive the abundance ratios among the isotopic species from the $J = 5-4$



The Astrophysical Journaltransitions observed at high S/N ratios by using the common excitation temperature for all isotopic species. Here we assume no anomalous excitation among the isotopic species. The HCCCN lines are not appropriate for obtaining a best estimate of the common excitation temperature because of the high opacity; therefore, we employ the HCC$^{13}$CN lines, which are the brightest among the three $^{13}$C species. As mentioned above, the column density and excitation temperature of HCC$^{13}$CN are obtained to be $(3.93 \pm 0.17) \times 10^{11}$ cm$^{-2}$ and $12.1 \pm 0.7$ K, respectively, where the errors denote one standard deviation (Table 2). The root-mean-square (rms) residual of the fit for the three HCC$^{13}$CN lines is 3.1 mK, which is roughly comparable with the noise level (rms) of the spectra (see Table 1). Thus, we use the excitation temperature of 12.1 K for all isotopic species as the common excitation temperature. The derived column densities are summarized in Table 3. The error of each column density, except for HCC$^{13}$CN, is evaluated from the error of the temperature of HCC$^{13}$CN and that of the integrated line intensity of each species, as each column density is derived by relative comparison with the intensity of HCC$^{13}$CN.

To evaluate the column densities of HCCCN and DCCCN, Sakai et al. (2009b) considered different excitation temperatures for the following cases: the "low" $J$ transition case for the $J = 5$–4 and 10–9 lines, and the "high" $J$ transition case for the $J = 10$–9 and 17–16 lines. They reported excitation temperatures of 9.7 K and 16.9 K, respectively, for the low and high $J$ transition cases. The temperature of $12.1 \pm 0.7$ K of HCC$^{13}$CN obtained in the present observations is just between these and is close to that of the "low" $J$ transition case.

The column density of HCCCN is derived from the ratio of the integrated intensities between HCCCN and HCC$^{13}$CN using the common excitation temperature mentioned above. However, the main hyperfine components of HCCCN are not suitable for this purpose because the optical depth is moderate (0.44) for the $J = 5$–4 line. Hence, we employ the integrated intensities of the two weak $F = 5$–5 and 4–4 components of $J = 5$–4 in order to derive the unsplitted intensity summing up all of the hyperfine components in the case of the optically thin condition. The unsplitted integrated intensity is thus evaluated to be $4.364 \pm 0.050$ K km s$^{-1}$, as listed in Table 1. On the other hand, the integrated intensity of HCC$^{13}$CN is $0.0676 \pm 0.0008$ K km s$^{-1}$. The rotational constants of HCCCN (4549 MHz) and HCC$^{13}$CN (4530 MHz) are very close to each other (Thorwirth *et al*. 2001), and hence the partition functions are comparable with each other for both species. Thus, the column density of HCCCN is estimated to be $2.52 \times 10^{13}$ cm$^{-2}$ from the column density of HCC$^{13}$CN and the ratio of the integrated intensities between HCCCN and HCC$^{13}$CN. This column density is consistent with that of $(1.9 \pm 0.5) \times 10^{13}$ cm$^{-2}$ obtained by the fitting of the three transitions of $J = 5$–4, 10–9, and 12–11 (Table 2), and agrees with that of $(2.7 \pm 0.2) \times 10^{13}$ cm$^{-2}$ reported by Sakai et al. (2009b) for the "low" $J$ transition case.

- 6 -



### *3.2. Isotopic Ratios in HC$_3$N*

In this study, the column densities of the normal and isotopic species are derived with high precision from the intensities of the $J = 5$–4 transitions (Figure 1), which were measured with very high S/N ratios. We derive the $^{13}$C abundance ratios as [H$^{13}$CCCN]:[HC$^{13}$CCN]:[HCC$^{13}$CN]:[HCCCN] = 1.00:1.01 ± 0.02:1.35 ± 0.03:86.4 ± 1.6 in L1527, in which the errors denote one standard deviation derived from the rms noise of the spectra. Our observational results are mostly consistent with those of Taniguchi et al. (2016a), but our results are much more precise. In particular, the equality of the H$^{13}$CCCN and HC$^{13}$CCN abundances is now confirmed within the one-sigma confidence level of 2%. This is much more stringent than the result for TMC-1 CP (Takano et al. 1998). Furthermore, the [H$^{13}$CCCN]:[HC$^{13}$CCN]:[HCC$^{13}$CN] ratios obtained for L1527 are very close to those in TMC-1 CP.

In addition, The column densities of DCCCN and HCCC$^{15}$N are also derived from the intensities of the $J = 5$–4 transitions using the method used for the $^{13}$C species. Then the ratios of [HCCCN]/[HCCC$^{15}$N] and [DCCCN]/[HCCCN] are obtained to be 338 ± 12 and 0.0370 ± 0.0007 (1$\sigma$), respectively.

In general, calibration errors, weather conditions, and pointing differences can affect integrated intensities; however, these effects were minimized in this study because all of the GBT data were obtained simultaneously. Moreover, none of the assumptions regarding the LTE condition, the excitation temperature, or the beam filling factor cause systematic errors in the obtained ratios because they are nearly canceled out in taking the ratios. The results are summarized in Table 3.

### *3.3. Isotopic Ratio in HC$_5$N*

The $J = 17$–16 transition of H$^{13}$CCCCCN was observed at 44.09 GHz using the GBT, as shown in Figure 3. The normal species of HC$_5$N was observed in L1527 by Sakai et al. (2009b), and the column density and the excitation temperature are derived to be $(6.8 ± 1.4) \times 10^{12}$ cm$^{-2}$ and 14.7 ± 5.3 K, respectively, by using the $J = 16$–15, 17–16, and 32–31 lines. The column density of H$^{13}$CCCCCN is calculated to be $(7.2 ± 0.8) \times 10^{10}$ cm$^{-2}$ under the assumption of LTE at the excitation temperature of 14.7 K. Based on this, the isotopic ratio of [HCCCCCN]/[H$^{13}$CCCCCN] is evaluated to be 94 ± 29 (1$\sigma$), which is consistent with the [HCCCN]/[H$^{13}$CCCN] ratio of 86.4 ± 2.2 and the [HCCCN]/[HC$^{13}$CCN] ratio of 85.4 ± 2.4 listed in Table 4.





## 4. DISCUSSION

### *4.1. Anomaly of the $^{12}C/^{13}C$ ratios in $HC_3N$*

In this study, we unambiguously find that the $^{12}C/^{13}C$ ratios for the three carbon-atoms of $HC_3N$ vary in L1527. A similar anomaly is reported for $HC_3N$, CCS, $C_3S$, CCH, and $C_4H$ in TMC-1 CP and for CCH and c-$C_3H_2$ in L1527. Two possibilities are proposed as the origin of this anomaly. One is that the anomaly reflects a production pathway of molecules, as discussed in previous papers (Takano et al. 1998; Sakai et al. 2007, 2010b, 2013; Yoshida et al. 2015). If a parent molecule has equivalent carbon atoms by symmetry, it cannot produce different $^{12}C/^{13}C$ ratios in a particular part of a product molecule. For instance, the different $^{12}C/^{13}C$ ratios for the two carbon atoms in CCH would exclude formation from the electron recombination reaction of $C_2H_2^+$, which has equivalent carbon atoms, as a main formation pathway. Another possibility is that an exchange reaction is stabilizing the energetically stable isotopic species, for instance:

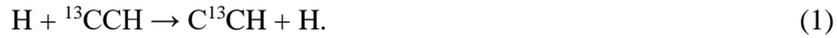

$$H + {}^{13}CCH \rightarrow C^{13}CH + H. \tag{1}$$

Because $C^{13}CH$ is more energetically stable than $^{13}CCH$ by 8.1 K (Sakai et al. 2010b), the $C^{13}CH/^{13}CCH$ ratio would be expected to be enhanced. Such exchange reactions after the formation of the molecules could contribute to the anomaly.

In the case of $HC_3N$, Takano et al. (1998) proposed the former possibility to explain the isotope anomaly in TMC-1 CP. $HC_3N$ is primarily formed through the reaction of $C_2H_2$ + CN, and the $^{12}C/^{13}C$ ratios of $C_2H_2$ and CN can differ as a result of their different formation histories and $^{13}C$ fractionation processes. For instance, the $^{12}C/^{13}C$ ratio in CN can be **lower** for the following fractionation reaction:

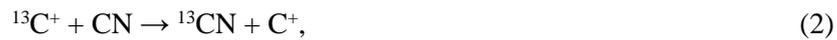

$$^{13}C^+ + CN \rightarrow {}^{13}CN + C^+, \tag{2}$$

which is exothermic by 31.1 K (Roueff et al. 2015). On the other hand, a similar reaction does not occur for $C_2H_2$. If the CN part of $HC_3N$ does come from the CN molecule, the $^{12}C/^{13}C$ ratio for the nitrile carbon could be lower than those for the other two carbon atoms coming from $C_2H_2$. This scheme is also possible in L1527 if $HC_3N$ was produced in a cold starless phase before the onset of star formation.

On the other hand, the second possibility does not significantly contribute to the $^{12}C/^{13}C$ anomaly in $HC_3N$. Because $HC_3N$ is a closed-shell molecule, exchange reactions similar to Equation (2) seem unlikely. This conclusion can be further confirmed by the following consideration. The zero-point vibrational energies of $H^{13}CCCN$, $HC^{13}CCN$, and $HCC^{13}CN$ are lower than that of the normal species by 48.9, 56.8, and 63.8 K, respectively (Takano et al. 1998). If the exchange reactions are responsible for the isotopic ratios, the abundances of $H^{13}CCCN$,





HC$^{13}$CCN, and HCC$^{13}$CN should increase in that order. However, the measured abundance ratios, that is [H$^{13}$CCCN] ≈ [HC$^{13}$CCN] < [HCC$^{13}$CN], differ from this prediction. Indeed, the equilibrium ratio is 0.5:1.0:2.0 at 10 K and 0.7:1.0:1.3 at 25 K if the $^{12}$C/$^{13}$C anomaly is caused by some thermal isotope exchange reactions. In the range of the gas kinetic temperature (10 K for the cold starless phase and 25 K for the WCCC condition), the observed ratios cannot be explained by the thermal equilibrium. Therefore, the anomaly of the $^{12}$C/$^{13}$C ratios of HC$_3$N in L1527 likely reflects the HC$_3$N formation mechanisms.

The [H$^{13}$CCCN]:[HC$^{13}$CCN]:[HCC$^{13}$CN] ratios in L1527 are very similar to those of TMC-1 CP. This suggests that reaction (2) followed by the reaction C$_2$H$_2$ + CN could be at work in L1527 as well as in TMC-1 CP. Hence, it is likely that the HC$_3$N observed in L1527 was produced in a similar low temperature condition to that in TMC-1 CP, i.e., it could be a remnant of a cold starless phase of the cloud and therefore produced before the start of WCCC. However, it should be noted that not all of the HC$_3$N in L1527 would be a remnant of this phase. Multi-transition observations of the $J$ = 5–4, 10–9, and 17–16 transitions of HC$_3$N reported by Sakai et al. (2009b) suggest that two HC$_3$N components having high (16.9 K) and low (9.7 K) temperatures exist for HC$_3$N in this cloud. In contrast to TMC-1 CP, the high temperature component in L1527 would originate from the WCCC effect in the vicinity of the protostar. The excitation temperature of 12.1 K of HC$_3$N obtained in this work using the $J$ = 5–4, 9–8, and 10–9 transitions is close to that of the low temperature component, and the ratios are determined by using the $J$ = 5–4 transition. Hence, the present ratios are likely to be characteristic of the low temperature component. The ratios of the high temperature component are also interesting, and observations of the $^{13}$C species of HC$_3$N in high $J$ transitions in L1527 are awaited.

## 4.2. Dilution of $^{13}$C in HC$_3$N

It has been reported that the $^{13}$C atom is diluted in carbon-chain molecules. Here, we examine the $^{12}$C/$^{13}$C ratios of HC$_3$N in L1527 in this context. The $^{12}$C/$^{13}$C ratios for H$^{13}$CCCN, HC$^{13}$CCN, and HCC$^{13}$CN are determined to be 86.4 ± 2.2, 85.4 ± 2.4, and 64.2 ± 1.5, respectively. The HCCCN/HCC$^{13}$CN ratio (64) is almost comparable to the elemental ratio of 60–70 in the local interstellar medium (Lucas & Liszt 1998; Milam et al. 2005). On the other hand, the HCCCN/H$^{13}$CCCN and HCCCN/HC$^{13}$CCN ratios are significantly higher than the elemental ratio. The average $^{12}$C/$^{13}$C ratio of HC$_3$N, defined as

$$R_{av}(\text{HC}_3\text{N}) = \frac{3\,[\text{HC}_3\text{N}]}{[\text{H}^{13}\text{CCCN}] + [\text{HC}^{13}\text{CCN}] + [\text{HCC}^{13}\text{CN}]}$$

is 77 ± 4, which is slightly higher than the elemental ratio. This means that $^{13}$C is likely diluted in HC$_3$N. The average $^{12}$C/$^{13}$C ratios for CCH and c-C$_3$H$_3$ are reported to be > 200 (Sakai et al.

The Astrophysical Journal2010b) and 150 ± 30 (Yoshida et al. 2015), and hence, the dilution in $HC_3N$ is not as significant as in CCH or $c-C_3H_2$. This trend can also be seen in TMC-1 CP. According to Takano et al. (1998), the average $^{12}C/^{13}C$ ratio in TMC-1 CP is 69, which is almost within the range of the elemental ratio. By contrast, the average $^{12}C/^{13}C$ ratios for CCS, CCH, and $C_4H$ in TMC-1 CP are 87 (Sakai et al. 2013), > 200, and 105 (Sakai et al. 2013), respectively. Hence, the $^{13}C$ dilution is less significant in $HC_3N$ than in the other carbon-chain molecules mentioned above. The similarity between L1527 and TMC-1 CP in terms of $^{13}C$ dilution in $HC_3N$ further supports the assumption that the $HC_3N$ observed in L1527 was mainly formed under a similar condition to that in TMC-1 CP (i.e., a low temperature condition).

The average $^{12}C/^{13}C$ ratios for the longer cyanopolyynes $HC_5N$ and $HC_7N$ in TMC-1 CP are 94 ± 6 (Taniguchi et al. 2016b) and > 52 (Langston & Turner 2007), respectively. These ratios are not **very** different from the average $^{12}C/^{13}C$ ratio in $HC_3N$. This trend can be seen in L1527. Our observations of $H^{13}CCCCCN$ in L1527 using GBT reveal a $^{12}C/^{13}C$ ratio of 94 ± 29, which is comparable to the average $^{12}C/^{13}C$ ratio in $HC_3N$.

Above all, the dilution of $^{13}C$ in cyanopolyynes is less significant than in other series of carbon-chain molecules, e.g., CCH (Sakai et al. 2010b), in both TMC-1 CP and L1527. This contradicts the results of the chemical model simulation by Furuya et al. (2011), which predict nearly the same dilution both for $HC_3N$ and CCH. This result implies that the formation region of $HC_3N$ (and probably $HC_5N$) is different from that for CCH and $c-C_3H_2$. Indeed, it is well known that different series of carbon-chain molecules (i.e., $C_nH$, $C_nS$, and $HC_nN$) show different distributions in TMC-1 CP (Olano et al. 1988, Hirahara et al. 1992, Langer et al. 1995, Pratap et al. 1997, Dickens et al. 2001). It is important to delineate distributions of $HC_3N$ in the protostellar core of L1527 and to compare these with the CCH distribution (Sakai et al. 2010a).

*4.3. Isotopic Fractionation of $^{15}N$ and D in $HC_3N$*

The $[HCC^{13}CN]/[HCCC^{15}N]$ ratio in L1527 is found to be 5.26 ± 0.19 (1σ), resulting in a $[HCCCN]/[HCCC^{15}N]$ ratio of 338 ± 12. This is the first observation of the $^{14}N/^{15}N$ ratio of this molecule, and is close to the ratio of $[H^{13}CN]/[HC^{15}N] = 5.56 ± 0.92$ in TMC-1 CP reported by Ikeda et al. (2002).

Using the Genesis solar wind sampling measurements, the $^{14}N/^{15}N$ ratio was determined to be 441 ± 6 (Marty et al. 2011), which is considered to be indicative of the composition of the so-called "proto-solar nebula" (PSN), from which our Sun was formed. In molecular clouds, the following $^{14}N/^{15}N$ ratios were measured from abundant N-bearing species. Observations of $NH_3$ reveal comparable ratios to those found in the PSN (e.g., Gerin et al. 2009, Lis et al. 2010, Daniel

- 10 -

et al. 2013), while $N_2H^+$ shows a ratio of ~1000 ± 200 in the prestellar core L1544 (e.g., Bizzocchi et al. 2013). HCN and HNC have lower ratios: 140–360 (Hily-Blant et al. 2013), 160–290 (Wampfler et al. 2014), and 237 (−21, +27) (Lucas & List 1998). Observations of CN and HCN reveal that the ratio in the interstellar matter surrounding our Sun is as low as 290 ± 40 (Adande & Ziurys 2012). The $^{14}N/^{15}N$ ratio obtained from $HC_3N$ in this study is similar to the ratios obtained for CN, HCN and HNC, which commonly have a nitrile.

The ratio of [HCCCN]/[DCCCN] in L1527 is obtained to be 0.0370 ± 0.0010 (1σ), which agrees with the previously observed ratio of 0.031 ± 0.011 within the given error range (Sakai et al. 2009b). The ratio is higher by a factor of 2 than the ratio reported for TMC-1 CP (Turner 2001) and is roughly half of the ratio in the carbon-chain-rich prestellar core L1544 (Howe et al. 1994).

## 5. SUMMARY

1. The rotational transitions of the $^{13}C$, $^{15}N$, and D isotopic species and the normal species of $HC_3N$ in the low-mass star-forming region L1527 were observed using the GBT and the Nobeyama 45 m telescope. In particular, the $J = 5–4$ lines of the three $^{13}C$ isotopic species were observed with high S/N ratios.

2. The $^{13}C$ isotopic ratios (fractionation) of $HC_3N$ are accurately determined to be [$H^{13}CCCN$]:[$HC^{13}CCN$]:[$HCC^{13}CN$]:[HCCCN] = 1.00:1.01 ± 0.02:1.35 ± 0.03:86.4 ± 1.6 (1σ). Based on a consideration of the zero-point energies of each isotopic species, these ratios cannot be explained as the results of carbon isotope exchange reactions after the formation of $HC_3N$. The ratios obtained for L1527 are very similar to those reported for the starless cloud TMC-1 CP, and it is likely that the $HC_3N$ in L1527 is a remnant of a cold starless phase of this cloud.

3. The column densities of $H^{13}CCCN$ and $HC^{13}CCN$ in L1527 are found to be close to each other. Hence, the equality of the $^{12}C/^{13}C$ ratio for these two $^{13}C$ species is established unambiguously. This result puts a stringent constraint in the production pathway of $HC_3N$ and strongly supports the production scheme of $HC_3N$ *via* $C_2H_2$ and $C_2H_2^+$.

4. The ratios of [HCCCN]/[DCCCN] and [HCCCN]/[$HCCC^{15}N$] in L1527 are precisely obtained as 0.0370 ± 0.0007 and 338 ± 12, respectively. The ratio of [HCCCN]/[DCCCN] is consistent with the previously reported ratio. On the other hand, the ratio of $^{14}N/^{15}N$ *via* $HC_3N$ is likely lower than that of the proto-solar nebula.

**Acknowledgment**

We thank the staff at the Nobeyama Radio Observatory and the National Radio Astronomy

Observatory for help with the observations. We thank Dr. David Frayer for his help in the observations at the GBT. M.A. thanks Grant-in-Aid for Scientific Research on Innovative Areas (Grant No. 25108002), Grant-in-Aid for Scientific Research (C) (Grant No. 15K05395), and the Institute for Quantum Chemical Exploration.

The Astrophysical JournalTable 1. Molecular Lines Observed in L1527.

| Transition | Species | Frequency [a] (GHz) | Telescope | $T_{MB}$ (K) | $\Delta v$ [b] (km s$^{-1}$) | $W$ [c] (K km s$^{-1}$) | $W_{all}$ [d] (K km s$^{-1}$) | rms in $T_{MB}$ (mK) | $V_{LSR}$ (km s$^{-1}$) |
|---|---|---|---|---|---|---|---|---|---|
| $J = 5$–$4$ | H$^{13}$CCCN [e] | 44.0841622 [f] | GBT | 0.0703(5) | 0.633(5) | 0.0473(7) | 0.0486(7) | 1.6 | 5.9 |
| | HC$^{13}$CCN [e] | 45.2973345 [f] | GBT | 0.0745(6) | 0.630(6) | 0.0500(8) | 0.0513(9) | 2.1 | 5.9 |
| | HCC$^{13}$CN [e] | 45.3017069 [f] | GBT | 0.0937(5) | 0.660(4) | 0.0658(8) | 0.0676(8) | 2.2 | 5.9 |
| | HCCCN [e] | 45.4903138 [f] | GBT | 3.275(13) | 0.785(3) | 2.738(23) | 2.814(23) | 2.6 | 5.9 |
| | $F = 5$–$5$ | 45.4888386 [g] | GBT | 0.1380(9) | 0.394(3) | 0.0579(8) | 4.364(50) [h] | 2.6 | 5.92 |
| | $F = 4$–$4$ | 45.4921104 [g] | GBT | 0.1415(6) | 0.388(2) | 0.0585(5) | | 2.6 | 5.91 |
| | HCCC$^{15}$N | 44.1672678 [f] | GBT | 0.0265(4) | 0.444(8) | 0.0125(4) | ← [i] | 2.1 | 5.94 |
| | DCCCN [e] | 42.2155827 [f] | GBT | 0.2007(14) | 0.662(5) | 0.1413(21) | 0.1452(21) | 2.0 | 5.9 |
| $J = 9$–$8$ | HC$^{13}$CCN [e] | 81.5341106 [f] | NRO [j] | 0.039(8) | 1.00(25) [j] | 0.041(10) | 0.041(10) | 9.1 | 6.0 |
| | HCC$^{13}$CN [e] | 81.5419806 [f] | NRO [j] | 0.046(7) | 1.7(3) [j] | 0.084(16) | 0.085(16) | 10.2 | 5.9 |
| $J = 10$–$9$ | H$^{13}$CCCN [e] | 88.166832 [f] | NRO [k] | 0.083(8) | 0.43(5) | 0.038(8) | 0.039(8) | 14.4 | 6.1 |
| | HC$^{13}$CCN [e] | 90.593059 [f] | NRO [k] | 0.095(9) | 0.53(5) | 0.053(10) | 0.054(10) | 14.4 | 6.0 |
| | HCC$^{13}$CN [e] | 90.601777 [f] | NRO [k] | 0.120(7) | 0.52(4) | 0.066(8) | 0.067(8) | 14.5 | 5.9 |
| | HCCCN [e] | 90.979023 [f] | NRO [k] | 4.115(18) | 0.607(3) | 2.660(25) | 2.678(25) | 18.1 | 6.0 |
| $J = 12$–$11$ | HCCCN [e] | 109.1736340 [f] | NRO [k] | 4.79(3) | 0.622(5) | 3.17(4) | 3.19(4) | 28.3 | 5.9 |
| $J = 17$–$16$ | H$^{13}$CCCCCN [e] | 44.0864313 [f] | GBT | 0.0129(5) | 0.472(22) | 0.0065(3) | ← [i] | 2.1 | 5.89 |

**Note.** The numbers in parentheses are errors in units of the last significant digits.

[a] Rest frequency.

[b] FWHM obtained by Gaussian fit.

[c] $W = \int T_{MB} dv$.

[d] Estimated integrated intensities summing up all hyperfine components. For example, the relative intensities are $F = 5$–$5$:$6$–$5$:$5$–$4$:$4$–$3$:$4$–$4$ = 0.0133:0.3939:0.3200:0.2593:0.0133 for the $J = 5$–$4$ transitions (Townes & Schawlow 1955).

[e] Blended lines of three hyperfine components $\Delta F = +1$ for the $J = (J'' + 1) - J''$ rotational transitions.

[f] The Cologne Database for Molecular Spectroscopy (Müller et al. 2001, 2005). Laboratory data for HC$_3$N and DC$_3$N were reported by Thorwirth et al. (2001) and Spahn et al. (2008), respectively. For H$^{13}$CCCCCN, the molecular constants come from Bizzocchi et al. (2004).

[g] Lafferty & Lovas 1978.

[h] The unsplitted integrated intensity estimated from $\Delta F = 0$ components summing up all the hyperfine components of the $J = 5$–$4$ transition in the case of the optically thin condition.

[i] $W_{all}$ is equal to $W$ because of no hyperfine splittings.

[j] Wide band AOS: The line widths are overestimated.

[k] High-resolution AOS.

Table 2. Column densities $N$ and excitation temperatures $T$ of the normal and $^{13}$C isotopic species of HC$_3$N in L1527. [a]

| Species | Lines used | $N$ (cm$^{-2}$) [b] | $T$ (K) [b] |
|---|---|---|---|
| H$^{13}$CCCN | $J = 5$–4, 10–9 | $2.8 \times 10^{11}$ | 10 |
| HC$^{13}$CCN | $J = 5$–4, 9–8,[c] 10–9 | $(2.8 \pm 0.6) \times 10^{11}$ | $12 \pm 3$ |
| HCC$^{13}$CN | $J = 5$–4, 9–8,[c] 10–9 | $(3.93 \pm 0.17) \times 10^{11}$ | $12.1 \pm 0.7$ |
| HCCCN | $J = 5$–4,[d] 10–9, 12–11 | $(1.9 \pm 0.5) \times 10^{13}$ | $16 \pm 4$ |

[a] The dipole moment $\mu = 3.73$ D (DeLeon & Muenter 1985) is used.

[b] The errors are 1σ.

[c] Half weighted in the least square fitting because of the lower S/N.

[d] The integrated intensity of $W_{all} = 2.814(23)$ K km s$^{-1}$ in Table 1 is used.

Table 3. Column densities $N$ and isotopic abundance ratios for HC$_3$N in L1527 derived from the integrated intensities $W_{all}$ of the observed transitions by using the common excitation temperature of 12.1 K.

| Species | $J = 5$–4 | | | | $J = 10$–9 |
|---|---|---|---|---|---|
| | $N$ | Ratio [a] | | | Ratio [a] |
| | (cm$^{-2}$) | $^{13}$C | H | $^{15}$N | $^{13}$C |
| H$^{13}$CCCN | $(2.91 \pm 0.06) \times 10^{11}$ [b] | 1.0 | | | 1.0 |
| HC$^{13}$CCN | $(2.95 \pm 0.07) \times 10^{11}$ [b] | $1.01 \pm 0.02$ | | | $1.4 \pm 0.4$ |
| HCC$^{13}$CN | $(3.93 \pm 0.17) \times 10^{11}$ [c] | $1.35 \pm 0.03$ | | | $1.7 \pm 0.4$ |
| HCCC$^{15}$N | $(7.5 \pm 0.3) \times 10^{10}$ [b] | | | 1.0 | |
| DCCCN | $(9.3 \pm 0.2) \times 10^{11}$ [b, d] | | $0.0370 \pm 0.0007$ | | |
| HCCCN | $(2.52 \pm 0.05) \times 10^{13}$ [b, e] | $86.4 \pm 1.6$ | 1.0 | $338 \pm 12$ | |

**Note.** The column densities $N$ and the isotopic abundance ratios are derived from $W_{all}$ in Table 1.

[a] The errors (1σ) come from only the errors of $W_{all}$ in Table 1.

[b] The error (1σ) is derived from both the error of the temperature of HCC$^{13}$CN and that of the integrated line intensity of each species. The former is canceled out by taking the ratios.

[c] The error (1σ) is obtained by least square fitting of the line intensities of the $J = 5$–4, 9–8, and 10–9 transitions, as shown in Figure 1.

[d] The dipole moment of DCCCN is $\mu = 3.74$ D (Coveliers et al. 1990), while that of the others are $\mu = 3.73$ D (DeLeon & Muenter 1985).

[e] The integrated intensity $W_{all}$ derived from the intensities of $F = 5$–5 and 4–4 is used.

Table 4. The $^{13}$C isotopic ratios of HC$_3$N and CCH in L1527 and TMC-1 CP.

| Ratio | L1527 | TMC-1 CP [a] |
|---|---|---|
| [HCCCN]/[H$^{13}$CCCN] | $86.4 \pm 1.6$ [b] | $79 \pm 11$ [c] |
| [HCCCN]/[HC$^{13}$CCN] | $85.4 \pm 1.7$ [b] | $75 \pm 10$ [c] |
| [HCCCN]/[HCC$^{13}$CN] | $64.2 \pm 1.1$ [b] | $55 \pm 7$ [c] |
| [CCH]/[C$^{13}$CH] | $\geq 80$ [d] | $\geq 170$ [d] |
| [CCH]/[$^{13}$CCH] | $\geq 135$ [d] | $\geq 250$ [d] |

[a] The cyanopolyyne peak (CP).

[b] The errors come from the errors of $W_{all}$ in Table 1.

[c] Takano et al. 1998.    [d] Sakai et al. 2010b.

**FIGURES**

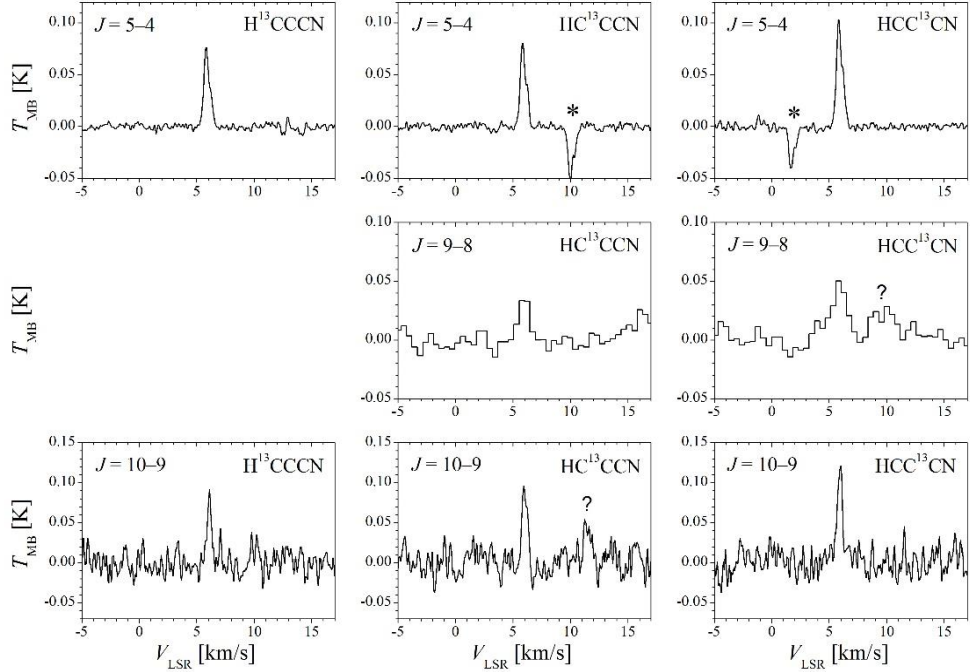

Figure 1. Observed lines of the $J = 5$–$4$, $9$–$8$, and $10$–$9$ transitions of the isotopic species $H^{13}CCCN$, $HC^{13}CCN$, and $HCC^{13}CN$ in L1527. The $J = 5$–$4$ transitions were observed with the Green Bank 100 m telescope. The $J = 9$–$8$ and $10$–$9$ transitions were with the NRO 45 m telescope using the wide-band and high-resolution acousto-optical spectrometers, respectively. Negative features indicated by the asterisks are frequency switch artifacts of other lines. Features labeled by the question marks are artifacts.

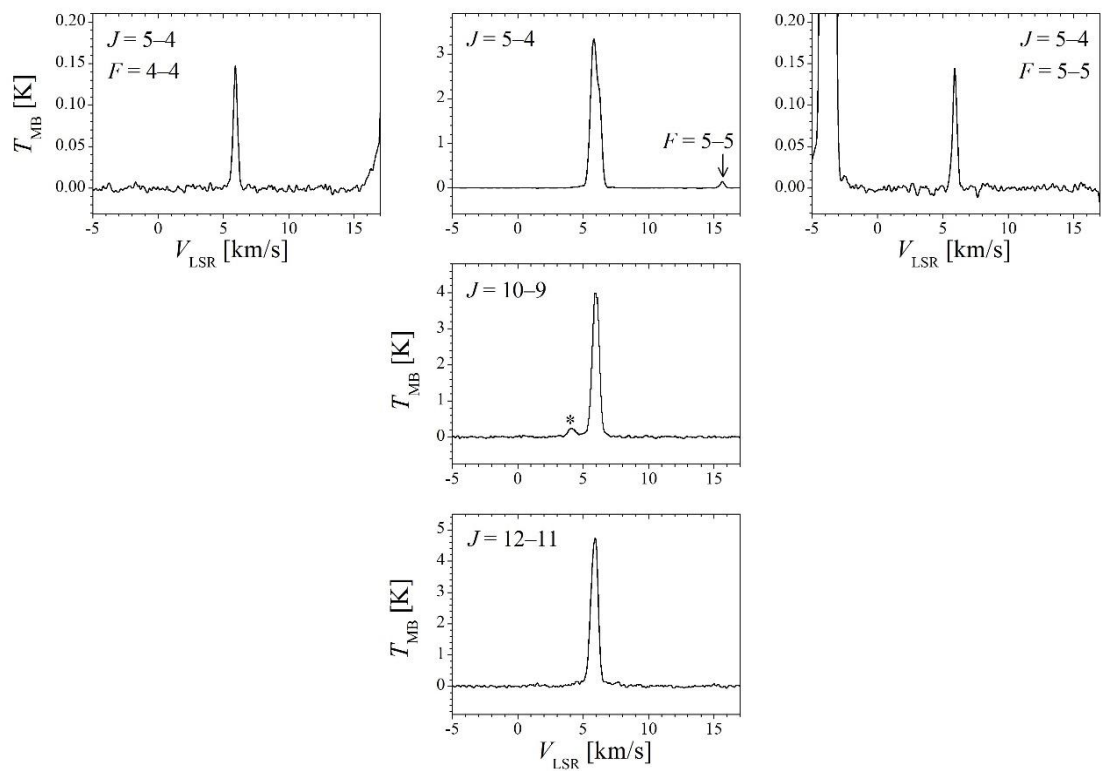

Figure 2. Observed lines of the $J$ = 5–4, 10–9, and 12–11 transitions of the normal species HCCCN in L1527. The blended lines of the three strong hyperfine components are shown in the middle column. The small feature indicated by the asterisk is an image-side band artifact.

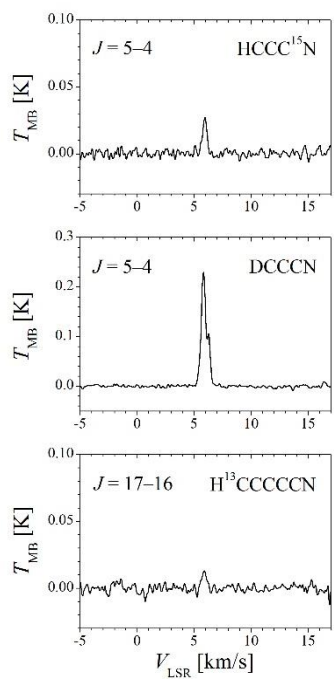

Figure 3. Observed lines of HCCC$^{15}$N, DCCCN, and H$^{13}$CCCCCN in L1527.